\documentclass{aa}  
\usepackage{natbib}
\usepackage{bm}
\usepackage{adjustbox}
\usepackage{blindtext}
\usepackage{ulem}
\usepackage{color}
\usepackage{threeparttable}
\usepackage{gensymb}
\usepackage{graphicx}

\usepackage[switch]{lineno}
\usepackage{txfonts}

\usepackage[colorlinks,
        linkcolor=blue,
           citecolor=blue]{hyperref}

\def\nh3{$\rm{NH_3}$}
\def\NH3{$\rm{NH_3}$}

\def\msun{\,$M_\odot$}

\def\lsun{\,$L_\odot$}

\def\um{\,$\mu\mathrm{m}$}

\def\kms{\,km~s$^{-1}$}
\def\cm2{\,$\rm{cm^{-2}}$}
\def\cm3{\,$\rm{cm^{-3}}$}

\newcommand{\ghz}{\,GHz}
\newcommand{\mhz}{\,MHz}

\usepackage{xcolor}

\usepackage{soul}

\begin{document} 

\title{Highly structured turbulence in high-mass star formation: an evolved infrared dark cloud G35.20-0.74 N}






   \author{Chao Wang
          \inst{1}
          \and
          Ke Wang\inst{2}
          \fnmsep\thanks{Corresponding author.}
          }

   \institute{Department of Astronomy, School of Physics, Peking University, 5 Yiheyuan Road, Haidian District, Beijing 100871, China
         \and
             Kavli Institute for Astronomy and Astrophysics, Peking University, 5 Yiheyuan Road, Haidian District, Beijing 100871, China\\
             \email{kwang.astro@pku.edu.cn}
             }

   \date{Received July 15, 2022; accepted xx xx, 2022}

 
  \abstract
{Massive stars are generally believed to form in supersonic turbulent environment. However, recent observations have challenged this traditional view. High spatial and spectral resolution observations of the Orion Molecular Cloud (OMC, the closest massive star formation region) and an infrared dark cloud (IRDC) G35.39 (a typical distant massive star formation region) show a resolution-dependent turbulence, and that high-mass stars are forming exclusively in subsonic to transonic cores in those clouds. These studies demand a re-evaluation of the role of the turbulence in massive star formation.}
{We aim to study the turbulence in a typical massive star-forming region G35.20-0.74 N (G35.20 in short) with a sufficient spatial resolution to resolve the thermal Jeans length, and a spectral resolution to resolve the thermal linewidth.}
{We use the Atacama Large Millimeter/submillimeter Array (ALMA) dust continuum emission to resolve fragmentation, the Karl G. Jansky Very Large Array (JVLA) 1.2 cm continuum to trace ionized gas, and JVLA \nh3\ (1,1) to (7,7) inversion transition lines to trace linewidth, temperature, and dynamics. We fit those lines and remove line broadening due to channel width, thermal pressure, and velocity gradient to obtain a clean map of intrinsic turbulence.}
{We find that (1) the turbulence in G35.20 is overall supersonic, with mean and median Mach numbers 3.7 and 2.8, respectively. 
(2) Mach number decreases from 6-7 at 0.1~pc scale to $<$3 towards the central cores at 0.01~pc scale.
(3) The central ALMA cores appear to be decoupled form the host filament, evident by an opposite velocity gradient and significantly reduced turbulence.
Because of intense star formation activities in G35.20 (as compared to the relatively young and quiescent IRDC G35.39), the supersonic turbulence is likely replenished by protostellar outflows.
G35.20 is, thus, representative of an evolved form of IRDC G35.39.
More observations of a sample of IRDCs are highly demanded to further investigate the role of turbulence in initial conditions for the massive star formation.}
   {}

  \titlerunning{Highly Structured Turbulence in G35.20-0.74}
  \authorrunning{Wang \& Wang}

   \maketitle
%

\section{Introduction} \label{sec:1}

The formation of high-mass stars ($M_\star >8$\msun) remains a fundamental problem in modern astrophysics. Traditionally, high-mass stars are believed to form in highly turbulent gas \citep[e.g.,][]{Larson1981,  2012ApJ...745...61L, 2013MNRAS.432.3288S, 2013ApJ...779...96T}. Supersonic turbulence is important to provide the required support against gravitational collapse \citep{2003ApJ...585..850M}, by maintaining an equivalent ``turbulent Jeans mass'' that is {generally} much higher than the thermal Jeans mass (typically on order of 1\msun), allowing the formation of massive stars \citep[e.g.,][]{2011ApJ...735...64W,WangHierarchical}.
Observations of giant molecular clouds and massive {star-forming regions often reveal} supersonic turbulence. Recent examples include interferometric observations of \nh3\ in a sample of high-mass star formation regions \citep{2013MNRAS.432.3288S,Lu_2014, Lu_2018}.

However, it is important to note that those \nh3\ studies, and most {of the} similar observations toward massive {star-forming} regions in the literature, do not actually provide sufficient resolving power to properly study turbulence that are important for the initial conditions of star formation.
Owing to large distances to typical massive {star-forming} regions, observations must meet the 
 following minimum requirements: 
(1) a high spectral resolution that is sufficient to resolve thermal linewidths (sound speed 0.23\kms\ at 15 K); (2) a high spatial resolution that is sufficient to resolve thermal Jeans length (0.09~pc for a clump of $10^3$\msun\ in 1\,pc diameter, or $3''.5$ at a distance of 5 kpc); 
(3) a flux sensitivity that is sufficient
to resolve thermal Jeans mass (typically 1 $M_\odot$).

Notably, several VLA observations in NH$_3$ used a correlator setup with a 0.6-0.7 km ${\rm{s}^{-1}}$ channel width, especially for observations taken prior to the EVLA upgrade in 2010 \citep{Perley2011EVLAproject}.
In those data, the instrumental broadening is already a few times larger than the thermal broadening and thus cannot detect the thermal linewidth. 
A special marginal case is presented by \cite{WangKe2012}, who combined VLA-C configuration \nh3\ image cubes at 0.2\kms\ resolution with VLA-D configuration data at 0.6 \kms\ resolution to study clump P1 in the G28.34+0.06 infrared dark cloud (IRDC). 
They found a supersonic turbulence over the cloud-scale. However, all the dense cores showed a much reduced line-width in comparison to the clump-scale widths, which is indicative of turbulence dissipation from clump to core-scale.

Only a few studies achieved the aforementioned requirements \citep{2018A&A...611L...3S, 2018A&A...610A..77H, 2021RAA....21...24Y,2022ApJ...926..165L}, and surprisingly, all of them found that subsonic to transonic turbulence, instead of supersonic turbulence, are dominant. 
\cite{2018A&A...611L...3S} imaged IRDC G035.39-00.33 (G35.39 hereafter) in \nh3\ (1,1) and (2,2) lines at a spectral resolution of 0.2\kms\ down to scales of 0.07 pc. 
The star-forming cores are exclusively located in subsonic regions of the IRDC, and the entire IRDC is dominated by subsonic turbulence.
Similar results are observed in the Orion molecular cloud, the nearest high-mass star-forming region \citep{2018A&A...610A..77H, 2021RAA....21...24Y}, and in NGC 6334S \citep{2022ApJ...926..165L}. In particular, \cite{2021RAA....21...24Y} showed that the observed linewidth is resolution-dependent, changing from transonic to subsonic as the spatial resolution increases from $10^4$ AU to $10^3$ AU. 

These studies, reported by four independent groups using different analyzing methods, suggest that previous similar observations with insufficient spatial and/or spectral resolution may lead turbulence to {\it appear} resolution-dependent, and that massive stars are forming in low turbulence environments in OMC, G35.39 and NGC6334S. Therefore, it is of great interest to study turbulence in massive star-forming regions with both sufficient spatial and spectral resolution, to further investigate whether supersonic turbulence is necessary to form high-mass stars. 
In general a turbulence with low Mach number infers a significantly lower support from the turbulence than it was previously assumed, and magnetic fields may play a more important role to support against gravitational collapse in order to form high-mass stars.

In this context, here we present JVLA and ALMA observations of the massive star formation region G35.20-0.74 N (G35.20 hereafter). The high spatial and spectral resolution enabled an in-depth investigation of the turbulence.
The paper is organized as follows. 
Section \ref{sec:2} presents observations and data.
Section \ref{sec:3} gives an overview of G35.20 with previous and our new observations.
Section \ref{sec:4} describes fitting to \nh3\ lines and results.
Section \ref{sec:5} discusses implications of results in the context of star formation.
Section \ref{sec:6} summarises main findings.


\section{Observations and data}\label{sec:2}

\subsection{JVLA}
We observed G35.20-0.74 N (RA (J2000), Dec (J2000): 18:58:13.00, 01:40:36.00)
using JVLA in its D configuration at K band on 2013 May 23 (project ID: 13A-373, PI: Ke Wang). The WIDAR correlator \citep{Perley2011EVLAproject} was tuned to cover \nh3 inversion transition lines from (1,1) to (7,7) at approximately 23.7-25.7\ghz, with a spectral resolution of 15.625 kHz (about 0.2\kms). At the same time, 15 subbands of 128\mhz\ are used to cover the continuum.
The VLA primary beam at K band is about $2'$.

Standard bandpass, flux, and gain of the target observation was calibrated by observing 3C454.3, 3C48 (0137+331), and J1851+0035, respectively. Corrections for baseline and atmospheric opacity are performed before calibration in CASA 4.7.2 (Common Astronomy Software Applications, \citealt{2007ASPC..376..127M}). Image cubes of  \nh3 lines are made using the multi-scale CLEAN algorithm \citep{2008ISTSP...2..793C} in CASA task \textsc{tclean} with Briggs weighting and a robust parameter of 0.5. The synthesized beam in \nh3 (1,1) is $2.9''\times1.9''$, PA=67.9$^\circ $ with the rms noise at 2.8~mJy $\rm{beam}^{-1}$. Integrated flux images of \nh3 (1,1) to (7,7) lines are presented in Fig.\ref{11-77}. To achieve the same resolution from \nh3 (1,1) to (7,7), we smooth all image cubes to the beam of the \nh3 (1,1) cubes. The line fitting (Section. \ref{sec:nh3fit}) is performed on the smoothed image cubes to avoid any artefacts originated from different beam shapes.

Additionally, we made a continuum image by multi-frequency synthesis line-free channels in the 15$\times$128\mhz\ wide sub-bands centered on 24.7\ghz, corresponding to 1.2\,cm. We used uv-taper and robust weighting during cleaning to reach a synthesized beam of $2.6''\times1.3''$, PA=75$^\circ$. The image rms noise is 18 $\mu \rm{Jy\,beam}^{-1}$, reaching a dynamical range better than 230.

\subsection{ALMA}
\begin{figure*}[htb!]
    \centering
    \includegraphics[width=0.9\textwidth]{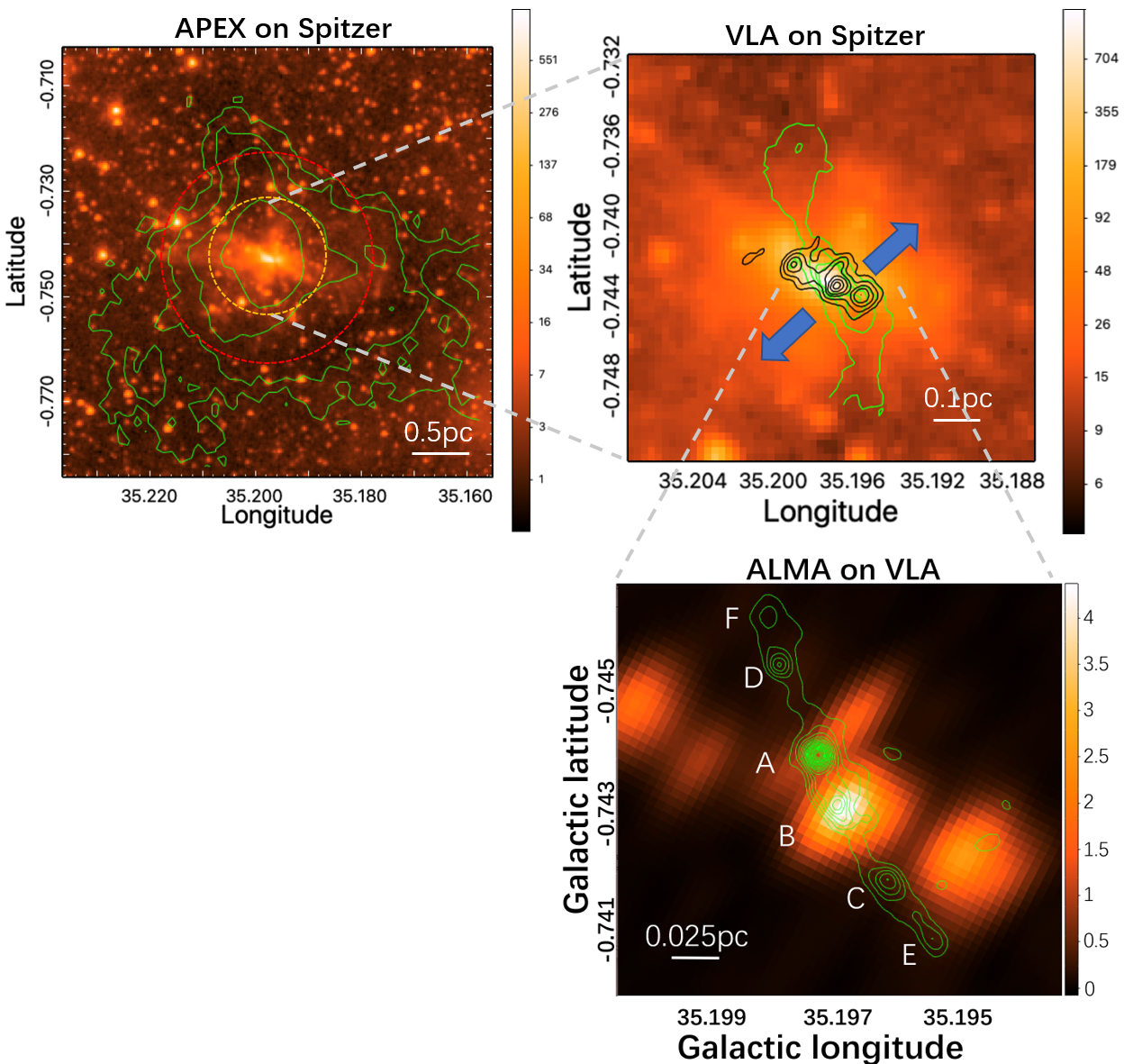}
    \caption{Overview of G35.20-0.74 N imaged by {\it Spitzer}, the Atacama Pathfinder Experiment (APEX), JVLA, and ALMA at various scales. \textbf{First row-left:} {\it Spitzer} 4.6\um\ emission \citep{2003PASP..115..953B} obtained by IRAC overlaid with a green contour map of the 870\um\ continuum emission from the APEX Telescope Large Area Survey of the Galaxy (ATLASGAL) survey \citep{2014A&A...565A..75C}. The unit of the color wedge is MJy $\rm{sr}^{-1}$ in IRAC map. The APEX 870\um\ contours start from 3 $\sigma$, then are 6, 12, 24 $\sigma$, where $\sigma = $90~mJy $\rm{beam}^{-1}$. The dashed yellow and red circles are of 1$'$ and 2$'$ diameter, with the latter being the VLA primary beam. \textbf{First row-right:} the same {\it Spitzer} 4.6\um\ emission overlaid with green contours showing the integrated intensity map of NH$_3$ (1,1), and black contours showing 1.2\,cm continuum emission, from our observations. The green contours start from 5 $\sigma$, then are 10, 15, 20 $\sigma$. Here 1 $\sigma$ equals to 1.2~mJy $\rm{beam}^{-1}$. The black contours start from 6 $\sigma$, then are 20, 48, 80, 130, 180, 230 $\sigma$. Here 1 $\sigma$ equals to 18~$\mu$Jy $\rm{beam}^{-1}$. Blue arrows show the direction of the CO outflow \citep{Birks2006}. \textbf{Second row:} the same VLA 1.2\,cm continuum in color scale overlaid with ALMA 870\um\ continuum emission in green contours. The color bar is in units of mJy $\rm{beam}^{-1}$. Green contours are from 9 to 230~mJy $\rm{beam}^{-1}$ with the step as 1.8~mJy $\rm{beam}^{-1}$. Cores A-F are marked following \cite{2014A&A...569A..11S}.}
    \label{over}
\end{figure*}

We retrieved the 870\um\ continuum image of G35.20 from the ALMA science archive (Project Code: 2011.0.00275.S, PI: Cesaroni, Riccardo). Details of the observations can be found in \cite{2013A&A...552L..10S}. The synthesized beam of the continuum image is $0.51''\times0.46''$, PA = 48$^ \circ$, with an rms noise level at 1.8~mJy $\rm{beam}^{-1}$. The ALMA primary beam at 350\ghz\ is about 16$''$, covering the central part of the VLA field, as marked in the middle panel of Fig.\ref{over}.


\section{G35.20 overview: dust, \nh3, and radio continuum in an evolved IRDC}\label{sec:3} 
Located at a parallax distance of 2.19 kpc \citep{2009ApJ...693..419Z}, the G35.20 molecular cloud is actively forming massive stars \citep{2014A&A...569A..11S}.
The dense clump coincides with a bright infrared source \object{IRAS 18556+0136}, and has been observed to have a high bolometric luminosity \citep[$>10^4$\lsun,][]{Gibb2003,2013A&A...552L..10S,ZhangYC2013},
molecular outflows \citep{1985MNRAS.217..217D,Gibb2003,Birks2006,Qiu2013},
radio continuum sources \citep[e.g.,][this work]{Gibb2003,Beltran2016}, 
and masers \citep[e.g.,][]{Brebner1987,Hutawarakorn1999,Surcis2012,Yan2013,Beltran2016}.

Fig. \ref{over} presents an overview of the G35.20 molecular cloud as observed by {\it Spitzer}, APEX, JVLA, and ALMA. 
The dense part as traced by APEX 870\um\ corresponds to $2.9\times 10^3$\msun\ in a radius of 0.87 pc \citep{2013A&A...552L..10S}.
Infrared extinction is visible in Fig. \ref{over} towards cloud extent (outermost ATLASGAL contours), spatially coincident with the {\it Spitzer} dark clouds identified in this region \citep{Peretto2009,Pari2020}.
However, the central part has developed into a butterfly shaped mid-IR nebula consistent with the active molecular outflow \citep{1985MNRAS.217..217D,2012ApJS..200....2L}. Hence, G35.20 represents an evolved IRDC.


Out of the apparently roundish clump seen in APEX 870\um\ image, the JVLA $\rm{NH}_3$ emission reveals a 0.5~pc long filament orientated north-south. The two ends of the filament slightly bend to opposite directions, resembling an integral shaped filament.
 The ALMA 870\um\ continuum image reveals an almost straight line filament harboring a string of six regularly spaced compact cores. Following \citet{2014A&A...569A..11S}, we label the ALMA cores as A to F in Fig. \ref{over}. 
Physical parameters of the cores are listed in Table \ref{tab:core}.
The central core B corresponds to an almost edge-on Keplerian disk and a possible binary system \citep{2013A&A...552L..10S,Beltran2016}.
Note that we have measured core fluxes above 3$\sigma$, while \cite{2014A&A...569A..11S} used a 5$\sigma$ cut, so the core fluxes reported here are slightly higher. We also used updated temperature values to compute the core masses reported in Tab. \ref{tab:core}. The masses are consistent with those listed in \cite{2014A&A...569A..11S}, which are in the range 0.5-11 ${M_\odot}$. We note that our mass estimates for cores A and B are likely overestimated since we have used lower temperatures compared to the ones measured in \cite{2014A&A...569A..11S} (see also Section. \ref{sec:tandn}). Besides, we also calculated cores' virial parameters and present them in Table. \ref{tab:core}. Details of the compute process are in Sec. \ref{sec:frag}.


The VLA 1.2\,cm continuum image reveals a lane of ionized gas misaligned with the \nh3\ and ALMA dust continuum emission. The angle between the ALMA dust filament and the VLA radio continuum lane is about 37\degree. The radio lane is consistent with VLA observations at 1.3\,cm and 2\,cm \citep{Heaton1988,Beltran2016}.
At this resolution, the 1.2\,cm radio continuum emission is condensed to at least five compact sources, which are further resolved at higher resolution by \cite{Beltran2016}. This lane spatially matches with the central part of the butterfly-shaped {\it Spitzer} nebula. \cite{Gibb2003} suggest that the radio lane traces a precising thermal radio jet, which was later confirmed by ALMA and VLBA observations \citep{2013A&A...552L..10S,Beltran2016}.

\begin{figure*}[htb!]
    \centering
    \includegraphics[width=1.0\textwidth]{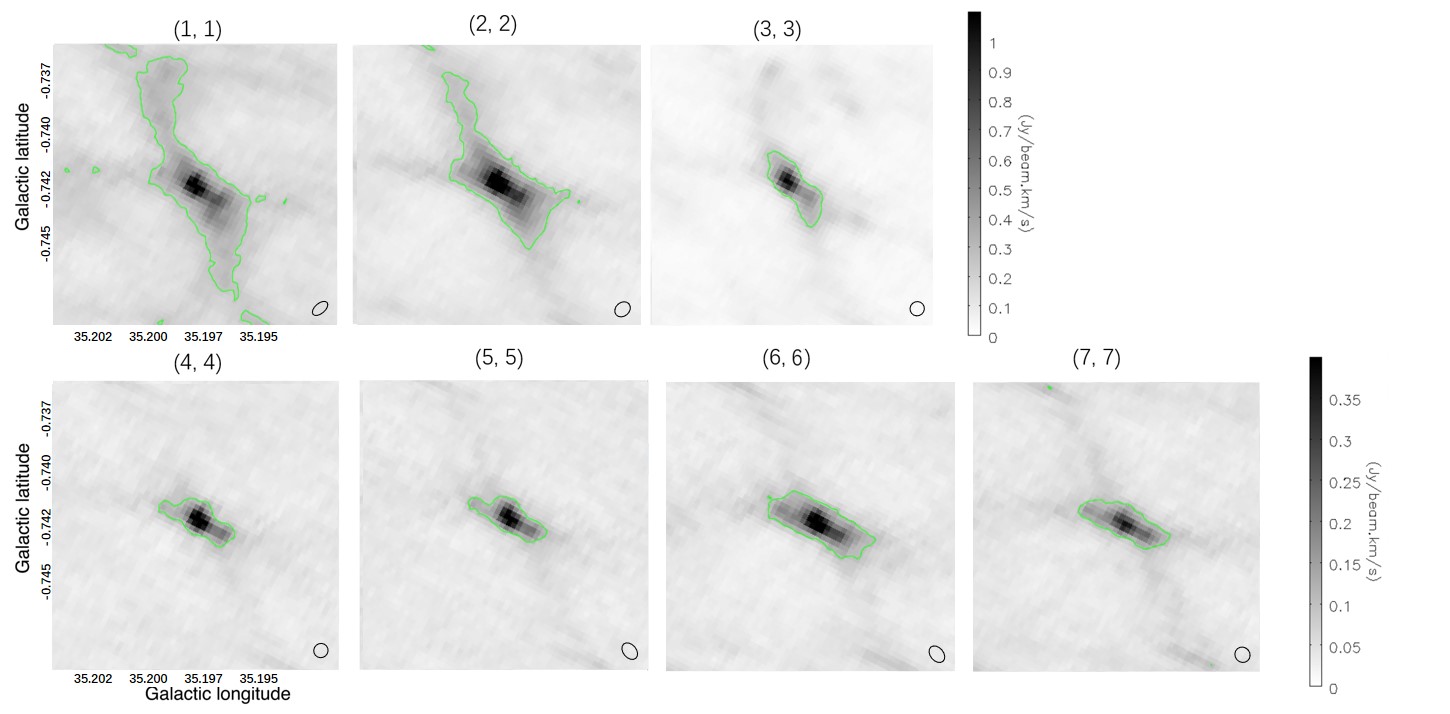}
    \caption{Flux integrated maps of JVLA NH$_3$ lines from (1,1) to (7,7) in G35.20-0.74 N. \textbf{Upper row:} from left to right are \nh3\ (1,1) to (3,3) shown in the same color bar. Ellipses in bottom-right corner are synthesized beams of each line image. Contours show the 3$\sigma$ level of each line integrated map. \textbf{Lower row:} similar to the upper row but for lines from (4,4) to (7,7), with a different color bar.}
    \label{11-77}
\end{figure*}

Fig. \ref{11-77} presents flux integrated maps of JVLA NH$_3$ from (1,1) to (7,7). The \nh3\ emission reveals an elongated filament of the size about 0.5$\times$0.1\,pc, an intermediate scale that bridges the larger spatial scale (clouds) traced in the APEX 870\um\ emission and the smaller scale (cores) traced in ALMA data. In the rest of the paper we refer this \nh3 feature as ``the filament/clump''.
\nh3 (1,1) and (2,2) reveal a larger extent of the elongated clump, while higher transitions from (3,3) to (7,7) highlight the central part, suggesting a warm central region surrounded by colder emission, which will be explored more in the next section. We also note that NH$_3$ (6,6) emission is stronger than those in the nearby (5,5) and (7,7) lines, likely due to the (6,6) transition being an ortho state. Comparing the ortho state (3,3) and (6,6) with other para states, we can find that the ortho/para ratio in NH$_3$ emission has a complex morphology. We will discuss the details in Section. \ref{sec:OPR}.

\section{\nh3\ fitting results}\label{sec:4}

\subsection{\nh3\ fitting} \label{sec:nh3fit}

\begin{figure*}[htb]
    \centering
    \includegraphics[width=1.0\textwidth]{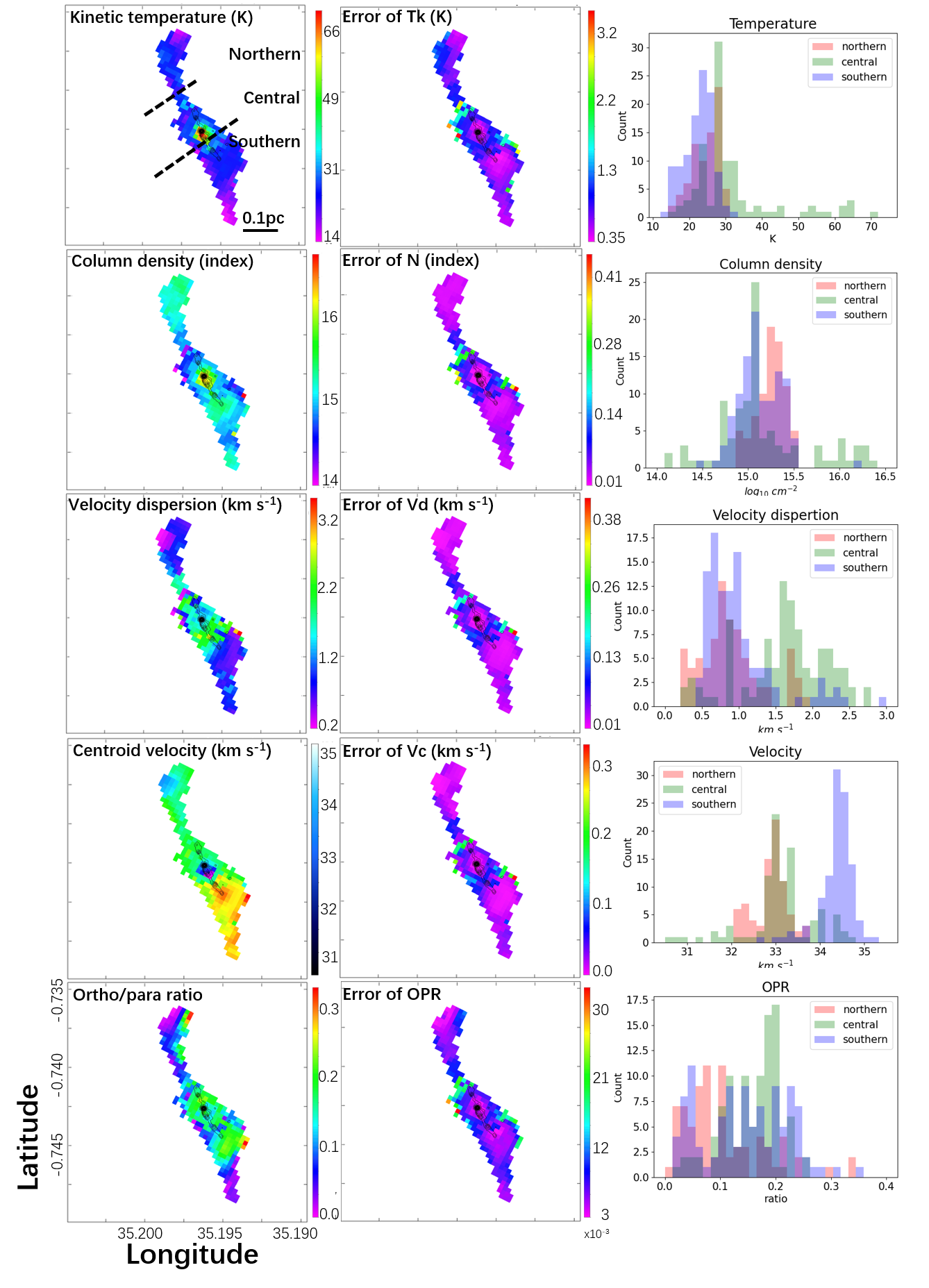}
    \caption{Maps of fitted parameters (left column) with errors (middle column) and histograms of pixel distribution of the Northern, Central, and Southern regions (right column). From top to bottom are the kinetic temperature, column density, velocity dispersion, centroid velocity, and OPR. The black contours is the same as that of the right panel in Fig.  \ref{over} which shows the ALMA 870\um\ emission. Scale bars on the right share the same units for parameter and fitting error.}
    \label{re}
\end{figure*}

We use python-based PySpecKit package \citep{2011ascl.soft09001G} to simultaneously fit the NH$_3$ (1,1) to (7,7) lines. The algorithm models the \nh3\ lines simultaneously using six fitting parameters, including excitation temperature (T$_{\rm ex}$), kinetic temperature (T$_k$), column density, ortho/para ratio (OPR), centroid velocity ($V_{\rm LSR}$) and velocity dispersion ($\sigma_v$). We do not use the cold-ammonia model presented in \cite{2012ApJS..200....2L} because of the relatively high temperature which is more than 50 K (in Fig. \ref{re}) toward the center of G35.20. 

Pixels with a signal-to-noise ratio (SNR) higher than 3 are included in our fitting. 
As shown in Fig. \ref{11-77}, the outer regions are only detected in \nh3 (1,1) and (2,2).
We tune the PyspecKit to fit lines in those regions by setting non-detected pixels as ``NAN'' so that the package will set upper limits automatically for all parameters ensuring the peak SNR of high-energy lines are less than 3. To check the influence of this limitation on both derived temperatures and column densities, we visually checked all fitted lines: none of the parameters are underestimated because of the limitation and based on the low residuals,  all fitted lines are reasonable. During the fitting, we check uncertainties and make sure they are under 10\%, otherwise we re-set the limitation and re-run the fitting.

In more detail, firstly we make a uniform parameter space to generate equally spaced parameters as the initial guess for the above model. We also adopt a number of criteria for parameters to ensure fitted results are reasonable: the kinetic temperature is from 3 K to $10^2$ K, the OPR is from 0 to infinite and the column density is ${10^{14} - 10^{17}}\ {\rm{cm}^{-2}}$. By limiting the range of parameters and using Monte Carlo Markov-Chain (MCMC) method, we obtain the best fitted result with the lowest relative residuals with the mean value at about 3\%. 


We have fitted the \nh3\ data cubes with multiple velocity components, and find that G35.20 is dominated by a single velocity component (detailed in appendix. \ref{sec:appendix}). Therefore, we adopt the fitting result from one velocity component throughout this paper. 
Fig. \ref{re} illustrates the fitting results, including kinetic temperature, column density, velocity dispersion, cetroid velocity, and OPR. Fitting error and histograms of pixel distributions of these parameters are also plotted.
The filamentary G35.20 cloud is divided into three regions based on their gas properties, the Northern, Central, and Southern, for easy description in the following text. 

\subsection{Temperature and column density: warm central and cool surroundings}\label{sec:tandn}
The fitted kinetic temperature of G35.20 is shown in the first row of Fig. \ref{re}. The temperature map is highly structured, featured by a warm central region up to 60-70 K and an almost uniform temperature of about 25 K across the entire clump, while towards the two ends temperature drops to $<$14 K. The coldest ends of the elongated clump have temperatures comparable to quiescent IRDCs \citep[e.g.][]{WangKe2012}, and may retain the initial conditions least affected by the star formation activities in the central region. The vast majority of pixels in G35.20 however, have higher temperatures than quiescent IRDCs, and are likely heated by the central region. We note that 70 K is clearly not the highest limit of the temperature in G35.20. \cite{2014A&A...569A..11S} used CH$_3$CN and CH$_3$OH, which trace denser gas than \nh3, and obtained up to 200 K towards the central cores.


The column density map reveals three distinct dense regions. Each region has a similar NH$_3$ column density at about ${10^{15.3}}\ {\rm{cm}^{ - 2}}$. The distribution of pixel values in those three regions is flat, with fitted uncertainty less than $10^{-2}$ dex. These three regions correspond to the North, Central, and Southern regions, respectively.


\subsection{Radial velocity and velocity dispersion: cores decoupled from filament}

The centroid velocity map clearly shows two gas streams, one at about 34.5\kms\ in the Southern region, and another at typically 33\kms in the Central and Northern regions (see Fig. \ref{re}). This is also obviously seen in the pixel histograms. Interestingly, towards the inner most part where ALMA cores A and B are located, the velocity shows distinct values of about 31-32\kms\ than the surroundings, suggesting these two cores may have been decoupled from the ambient gas in the Central region. Another similar trend is seen toward the southern end covered by the ALMA data, at core E, where the velocity is distinctly higher than the surroundings. Similarly, near the northern tip of the clump, a distinct velocity is also observed (blue color in the map), but that region is out of the ALMA field of view, so the presence of a dense dust continuum core is yet to be observed.

This northern tip corresponds to the lowest velocity dispersion in the entire G35.20 clump ($\sim 0.3$\kms), possibly representing the turbulence level of the natal clump unaffected by star formation. The Central region has higher velocity dispersion than the rest of the clump, typically 1.5-2\kms, compared to 0.6-0.9\kms\ in the Southern and Northern regions. Most of the ALMA cores coincide with reduced velocity dispersion compared to the surroundings, similar to the findings of \cite{WangKe2012}. This leads to a lower Mach number (Sect. \ref{sec:mach_num}) in the inner most Central region.


 
\subsection{\nh3 ortho/para ratio (OPR)} \label{sec:OPR}

The OPR map reveals a rich structure similar to that of the column density (see Fig. \ref{re}). The mean OPR across the clump is about 0.2, with higher values mainly concentrated in the Centarl and Southern regions. The Northern region and the two tips of the filament/clump feature the lowest OPR down to less than 0.1. We note that the Centarl/Southern regions harbor the ALMA cores, particularly cores A/B with active star formation, where heating and outflows may have increased the OPR as compared to the rest of the clump \citep[e.g.,][]{WangHierarchical}.  More detail on this point is discussed in Sect. \ref{sec:OPR_discuss}.


\subsection{Mach number: calm cores in turbulent gas} \label{sec:mach_num}
\begin{figure*}[!ht]
    \centering
    \includegraphics[width=1.0\linewidth]{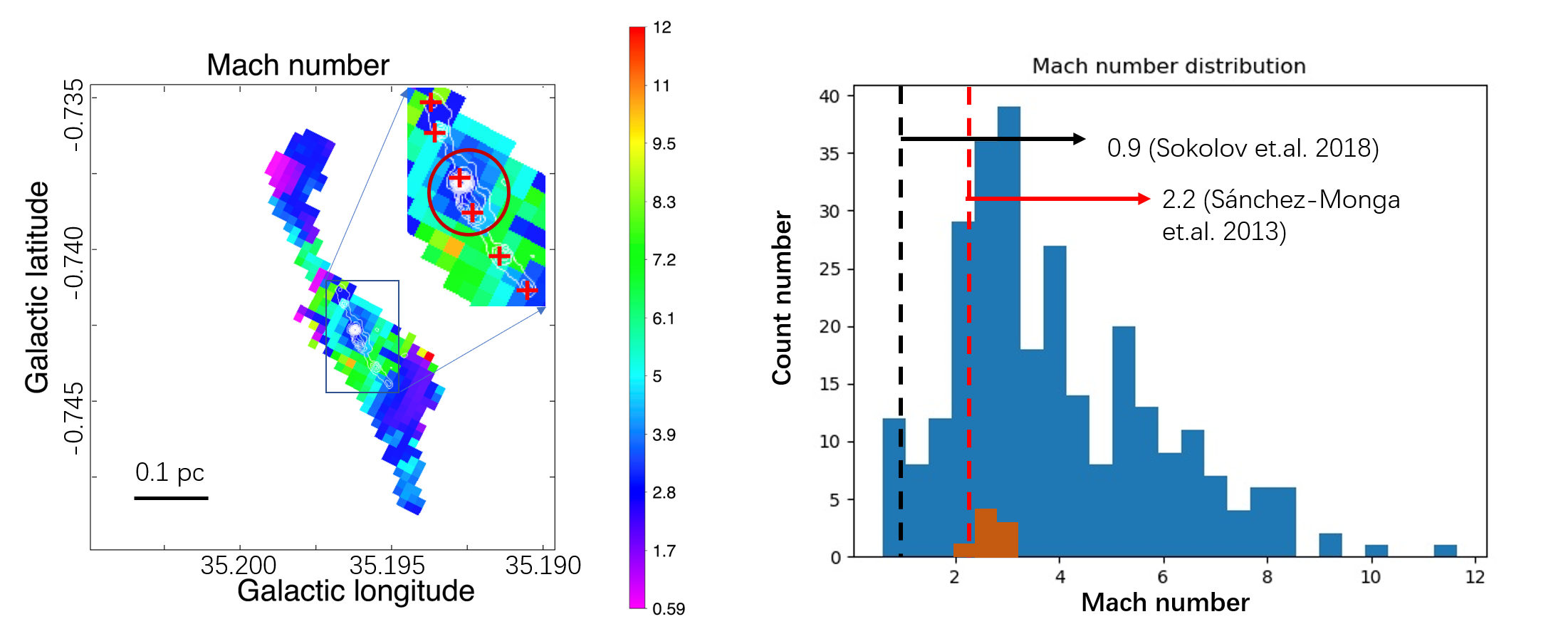}
    \caption{Mach number (left) and its pixel distribution (right).
    The white contours and red crosses mark the ALMA 870\um\ emission cores, with contour levels the same as in Fig. \ref{over}.
    The dashed black/red vertical lines mark the typical Mach number in the literature, representative of quiescent and protostellar stages, respectively. 
    The brown histogram corresponds to the brown circle of the central region (covering cores A/B), and the blue histogram is that of the whole G35.2 clump.}
    \label{mn}
\end{figure*}

To calculate the Mach number, we first correct the line width broadened by the effect from the finite channel width:

\begin{equation}
    {\sigma _{{V_{intrinsic}}}}^2 =\sigma _{V_{obs}}^2 -{\sigma _{{V_{channel}}}}^2
\end{equation}

\begin{equation}
    {\sigma _{{V_{channel}}}} = \frac{{0.2\;km \cdot {s^{ - 1}}}}{  {\sqrt {8\ln 2} }}
    \label{vch}
\end{equation}

Here the $\sigma _{V_{obs}}^2$ is derived from the fitted result (typically in the range 0.56-2.43 km ${\rm{s}^{-1}}$) and the $\sigma _{{V_{channel}}}^2$ is the channel width (in this work is 0.21 km ${\rm{s}^{-1}}$). The channel width contributes the 9.79\%$\pm$3.26\% of the intrinsic velocity dispersion. Then we calculate the non-thermal velocity dispersion from the fitted temperature and the $\sigma _{{V_{intrinsic}}}$ \citep{1983ApJ...270..105M}:

\begin{equation}
    {\sigma _{{V_{nt}}}}^2 ={\sigma _{{V_{intrinsic}}}}^2 - {\sigma _{{V_{th}}}}^2
\end{equation}

\begin{equation}
    {\sigma _{{V_{th}}}} =\sqrt{\frac{{{k_B}{T_{kin}}}}{{{m_{N{H_3}}}}}}
    \label{vth}
\end{equation}

The width caused by the temperature (thermal) 
is derived using Eq. \ref{vth}, is typically in the range 0.17-0.34 km ${\rm{s}^{-1}}$, and corresponds to 9.63\%$\pm$2.53\% of the intrinsic velocity dispersion. As the correction is based on each pixel which has a finite resolution, we need to deduct the velocity gradient from the large scale motion to get the line width purely caused by turbulence. This velocity gradient contributes the extra velocity dispersion at the pixel scale: the slight difference at two opposite edges of one pixel could make the velocity dispersion {\it appear} larger (see detailed methodology in \citet{Lu_2018}). Based on that, we derive the true non-thermal velocity dispersion using:

\begin{equation}
    {\sigma _{{V_{nt-true}}}}^2 ={\sigma _{{V_{nt}}}}^2 - {\sigma _{{V_{grad}}}}^2
\end{equation}

The large scale gradient is about 3.96 km ${\rm{s}^{-1}}$ ${\rm{pc}^{-1}}$, and contributes only about 2.85\%$\pm$1.79\% to the total velocity dispersion, a relatively smaller contribution compared to the other components. The reason is likely due to the fact that with our high-spatial resolution observations we have not detected large velocity gradients in the G35.20 region.

So far, we have carefully subtracted the line-width broadening originated from channel width, thermal motion, and velocity gradient, which respectively contribute to about 10\%, 10\%, and 3\% of the observed velocity dispersion. The rest of broadening is considered to be caused by turbulence. We then proceed to compute the sound speed based on the temperature and the Mach number making use of the true non-thermal velocity dispersion, using the equations:

\begin{equation}
    {c_s} = \sqrt{\frac{{{k_B}{T_{kin}}}}{{{m_p}{\mu_p}}}}
    \label{f1}
\end{equation}

\begin{equation}
    {M_{number}} = \frac{\sigma _{{V_{nt-true}}}}{c_s}
    \label{f2}
\end{equation}

Note that most previous interferometric \nh3\ observations had a larger channel width of 0.6 km ${\rm{s}^{-1}}$ (determined by the old VLA correlator), which was larger than the sound speed \citep[e.g.][]{2013MNRAS.432.3288S, Lu_2014, Lu_2018}. We stress the necessity of using a spectral resolution higher than that of thermal broadening, otherwise non-thermal motions would be overestimated. 




The Mach number map is shown in Fig. \ref{mn}. The mean value of the Mach number is about 3.7, and the pixel distribution peaks at 2.8. The error map is flat with a mean value of 0.016. The Mach number map is highly structured and similar to the velocity dispersion map: the two ends of the G35.20 have a lower Mach number (about 2) and the central region has a relatively higher value (about 6). Interestingly, towards the central cores the Mach number drops again to $<$3. 
We discuss this further in Section. \ref{discuss:turb}.

\begin{figure*}[htb!]
    \centering
    \includegraphics[width=1.0\textwidth]{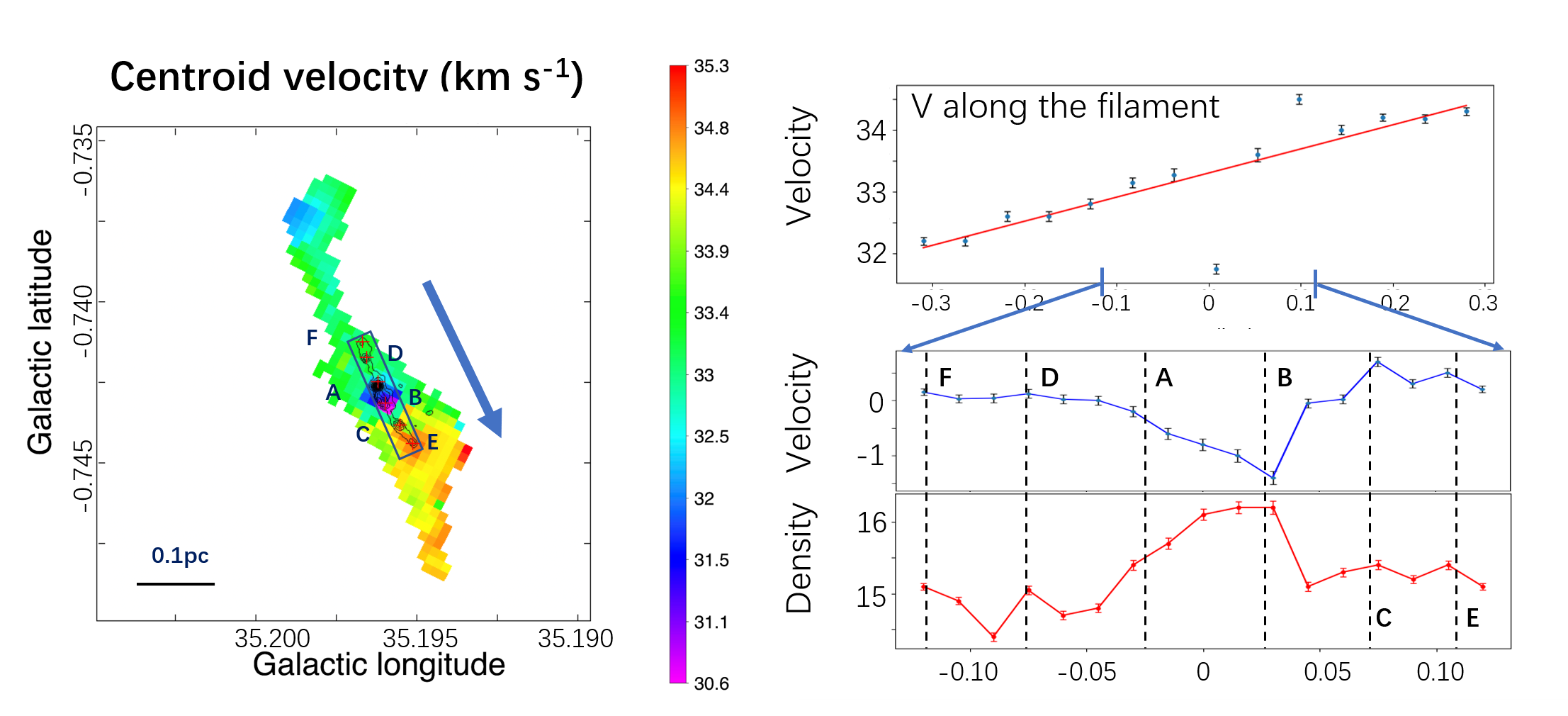}
\caption{
\textbf{Left:} \nh3\ fitted centroid velocity overlaid with ALMA 870\um\ contours as in Fig. \ref{re}. The box and arrow indicate how the PV cuts are made on the right panels.
\textbf{Right:}
The top panel is a PV cut showing the averaged centroid velocity (beam by beam) of the entire \nh3\ filament along the direction pointed by the arrow in the left panel, starting from the northern end down to the southern end of the \nh3\ filament shown in the left panel. The zero point locates between cores A and B. The red line is a linear fit excluding data between cores A and B.
The middle panel is a pixel-by-pixel (instead of beam by beam) PV cut along the black box region in the left panel, after subtracting a global velocity gradient (red line in top panel).
The bottom panel corresponds to the \nh3\ column density profile. Positions of the six ALMA cores A-F are marked with dashed vertical lines.
}
    \label{ve}
\end{figure*}

\subsection{Position-velocity diagram: wave-like pattern}
\label{sec: PV}
From the ALMA continuum map, the six cores are distributed almost regularly spaced along the filament. Separations among those cores are similar and the velocity of each core is slightly different from that of the host filament. Considering this, we make a position-velocity (PV) map (beam by beam) along the direction of the filament (the direction is shown in Fig. \ref{ve}) from the fitted velocity map to study the possible relationship between those cores and their host filament. The velocity is adopted as the average value of each sampled beam. To investigate the most active region in the central region, we also chose a rectangular region in the central part of G35.20 which corresponds to [-0.12, 0.12 pc] in the complete PV plot (0 point is between core A and B). This rectangular region covers all 6 cores and we plot the PV plot pixel by pixel, marking the location of cores from A to F. The result is shown in Fig. \ref{ve}. The six cores seem to belong to 3 pairs in the ALMA images: cores A and B seem connected, and similarly for C/E, and D/F. 

Interestingly, cores A/B have centorid velocities significantly different from the filament, suggesting that they may have been decoupled from the parental clump they are embedded in. This decoupling is also evident in the velocity dispersion and Mach number. The PV diagram in Fig. \ref{ve} clearly shows an opposite velocity gradient in cores A/B and in the host filament. The distinct velocity gradient in cores A/B is consistent with the kinematics traced by hot molecular lines \citep{2013A&A...552L..10S}.

As cores A and B have obvious different velocities with respect to the host filament, we masked pixels around these two cores within one beam size and fit the rest of the data in the PV map with a line. The linear fitting results in a velocity gradient of 3.96 km ${\rm{s}^{-1}}$ ${\rm{pc}^{-1}}$ along the \nh3\ clump. The fitting result is shown in Fig. \ref{ve}. This velocity gradient is typical of that measured at core scales \citep{2021RAA....21...24Y}, but it is $>$10 times larger than typical velocity gradients measured in longer filaments \citep{wangke2016FL, 2018RNAAS...2...52W, GeYF2022FL}.

After removing this global velocity gradient, the PV diagram shows some wave-like patterns. We then performed a Lomb-Scargle method to study the periodicity (method detailed in \citet{2018RNAAS...2...52W}). The most likely period is 0.025 pc (although other larger periods are also possible), consistent with the average separation of 0.023 pc between the ALMA cores \citep{2014A&A...569A..11S}.
This picture is consistent with the toy model proposed by \cite{2011A&A...533A..34H} to explain the formation of cores in one of the filaments of the L1517 star-forming region. However, the small number of data points in our observations limited us from a sophisticated comparison.

\section{Discussion}\label{sec:5}
\subsection{Fragmentation and stability} \label{sec:frag}

We use the APEX and ALMA 870\um\ dust continuum emission data to study the fragmentation of the G35.20 cloud.
The APEX 870\um\ thermal dust emission (Fig. \ref{over}) reveals a clump with a diameter of 0.87~pc, encompassing  119.1 Jy of integrated flux.
With an averaged dust temperature of 25.9 K from our fitted results, this flux corresponds to
 $4.94\times10^3{M_\odot}$. Assuming a spherical geometry, its volume density is ${1.11 \times {\rm{1}}{{\rm{0}}^{\rm{4}}}\ {\rm{cm}^{ - 3}}}$. Under the derived temperature and density, the thermal Jeans mass is about 11.2 ${M_\odot }$ and the Jeans length is 0.32~pc.
Here the temperature is the current value elevated due to heating from the active star formation in the clump center, so the Jeans values should be regarded as the current values as well, and so should not be compared to the observed ALMA cores.
For a meaningful comparison, we should estimate the Jeans mass/length of the initial G35.20 clump, which is obviously not directly observed now, but we can make reasonable estimation. We assume an initial temperature of 15 K, typical for quiescent IRDCs and in agreement with the currently observed kinetic temperature in the two ends of the G35.20 filament/clump which are least affected by the central star formation activities. We also assume that the averaged volume density does not change significantly since the initial stage. Then, for 15 K and ${1.11 \times {\rm{1}}{{\rm{0}}^{\rm{4}}}\ {\rm{cm}^{ - 3}}}$, the Jeans mass is 4.95\msun, and Jeans length is 0.24 pc. The Jeans mass is comparable to the ALMA cores, but the Jeans length is 10 times higher than the mean separation of between the cores (Table. \ref{tab:core}). 

It is possible that the cores, after initial fragmentation, have drifted toward the center of the gravitational potential well, now at the location of cores A and B. A useful reference is that for a 1\kms\ velocity, it takes $10^6$ yr to move 1 pc.
Assuming a relative velocity of 0.2\kms\ between the cores (cf. Fig. \ref{ve}), it takes $10^6$ yr to move 0.2 pc, a few times the free-fall time scale of the G35.20 clump ($3.5\times10^5$ yr). But the spatial distribution of those cores presents a well-aligned filamentray structure which indicates that the fragmentation most likely happened after the formation of the filament. This drift scenario has little possibility.
Another possibility is that the filamentary morphology has played a role in fragmentation. The regularly spaced ALMA cores resemble a cylinder fragmentation rather than Jeans fragmentation in homogeneous medium \citep[e.g.][]{2011ApJ...735...64W,WangHierarchical,wangke2016FL,GeYF2022FL}. However, it is difficult to infer the initial conditions of the gas cylinder that fragmented to the ALMA cores. High-resolution observations of quiescent IRDCs sensitive to multiple spatial scales would constitute the best tools to better understand the initial steps of the fragmentation process.

Caution has to be taken for a proper Jeans analysis for fragmentation \citep[cf. recent examples][]{Saha2022,XuFW2023,JiaoWY2023,ZhangSJ2023}. For example, in our above analysis an underlying assumption is that the observed ALMA cores are directly fragmented from the initial clump. However, hierarchical structure is common in molecular clouds, and it is possible that there is an intermediate structure between the pc-scale clump to the 0.01 pc scale ``cores'' (Table. \ref{tab:core}). It is evident that the six ALMA cores seem to belong to three pairs: A/B, C/E, D/F (Sect. \ref{sec: PV}). Each of these pairs might have been an original 0.1 pc scale structure that fragmented to form the cores. The density of such 0.1 pc scale structures would be larger, making the corresponding Jeans mass smaller. The current ALMA data, however, only covers the central 0.2 pc of the clump. A full mosaic of the clump combining 12m and ACA, and ideally also singled-dish data, would be best suited to study the hierarchical fragmentation \citep{2011ApJ...735...64W,WangHierarchical} in G35.20. That is beyond the scope of this paper, but is critically important to understand the initial conditions for massive star formation.

We investigate the stability of the clump in G35.20 by studying its virial properties. The virial parameter of the entire clump is $<$0.2, estimated based on the above clump mass/size and FWHM linewidth of 3.2\kms\ measured from Effelsberg \nh3 (1,1) \citep{Wienen2012}. A smaller FWHM of 2.6\kms\ is reported by \citet{Urquhart2011} based on GBT \nh3 (1,1). So, the G35.20 clump, even with the current turbulence level, is gravitationally unstable. Here we have not taken into account magnetic fields, which may provide additional support against gravity at a similar level as turbulence. However, at the initial stage the line-width may have been even smaller. We can therefore conclude that the virial parameter of the initial clump is much smaller than 1. Similarly, we calculate cores' virial parameters based on their fitted column density, size and velocity dispersion. The result is presented in Table. \ref{tab:core}. If we use the velocity dispersion which is measured by Effelsberg (1.4\kms\ ) or GBT (1.1\kms\ ) to calculate cores' virial parameters, the result is similar to that from JVLA (1.2\kms\ ). Their differences are in the range of the error. The current ALMA cores have virial parameters mostly $<2$. Without knowing and counting magnetic fields, 4 of the 6 cores are in rough agreement with the ``turbulent core'' scenario of being in virial equilibrium \citep{McKee2002}. However the core masses are far less than $10^2$\msun\ massive cores in the scenario, even if they have accreted more mass since the initial stage as evident by the observed outflows. Additionally, the unstable clump has highly fragmented to at least six cores, and are not in a monolithic structure.

\begin{table*}
    \centering
    \caption{Physical parameters of the ALMA cores} 
    \setlength{\tabcolsep}{5pt}
    \label{tab:core}
    \linespread{1.5}
    \begin{tabular}{lllllllllll}
    \hline    
    Name &Flux & M  & Size & $M_{vir}$& $\alpha$ & $N_{NH_3}$ & $T$  & $\sigma _{{V_{obs}}}$ & $\sigma _{{V_{nt-true}}}$  & Mach \\
    &Jy&${M_\odot }$&$10^{-3}$pc&${M_\odot }$& &$log_{10}~cm^{-2}$&K&$km~s^{-1}$&$km~s^{-1}$& \\
    \hline
    A&0.7$\pm$0.2&10.6$\pm$4.3&8.0$\pm$0.9&9.9$\pm$1.0&0.9$\pm$0.3&15.8$\pm$0.1&71.8$\pm$0.5&1.46$\pm$0.03&1.42$\pm$0.03&3.0$\pm$0.1\\
    B&0.4$\pm$0.1&3.7$\pm$1.6&4.8$\pm$0.5&8.0$\pm$0.9&2.1$\pm$0.6&16.2$\pm$0.1&58.3$\pm$0.6&1.69$\pm$0.02&1.68$\pm$0.02&2.9$\pm$0.1\\
    C&0.2$\pm$0.1&2.2$\pm$0.7&6.4$\pm$0.8&14.2$\pm$1.5&6.4$\pm$0.5&15.4$\pm$0.1&39.1$\pm$0.6&1.95$\pm$0.05&1.94$\pm$0.05&5.9$\pm$0.2\\
    D&0.2$\pm$0.1&2.3$\pm$0.7&5.7$\pm$0.6&8.6$\pm$0.9&3.7$\pm$1.4&15.1$\pm$0.1&28.6$\pm$0.7&1.62$\pm$0.05&1.60$\pm$0.05&5.0$\pm$0.2\\
    E&0.2$\pm$0.1&2.1$\pm$0.6&4.1$\pm$0.3&4.3$\pm$0.5&2.0$\pm$0.6&15.5$\pm$0.1&27.5$\pm$0.4&1.35$\pm$0.03&1.12$\pm$0.03&3.7$\pm$0.1\\
    F&0.3$\pm$0.1&3.0$\pm$1.2&4.0$\pm$0.3&1.8$\pm$0.3&0.6$\pm$0.1&14.9$\pm$0.1&28.8$\pm$0.7&0.87$\pm$0.04&0.85$\pm$0.04&2.6$\pm$0.1\\
    \hline
    \end{tabular}
    \begin{tablenotes}
    \footnotesize
    \item 
    Physical parameters of the ALMA cores, as measured from ALMA dust continuum and VLA \nh3. The columns are core name, 870\um\ flux, mass, core diameter, virial mass, virial parameter, \nh3\ column density, \nh3 fitted kinetic temperature, observed velocity dispersion ($FHMW = \sqrt {{\rm{8ln2}}} \sigma $), intrinsic velocity dispersion, and Mach number.
    \end{tablenotes}
\end{table*}

\subsection{Supersonic turbulence and its dissipation towards cores}
\label{discuss:turb}

The Mach number map in Fig. \ref{mn} provides a snapshot of the current status of the turbulence in G35.20. The spatial distribution of Mach number is highly structured. Overall, the entire clump is mostly supersonic, but the two ends are transonic to subsonic, with Mach number down to 2 or even lower than 1. Towards the central 0.15 pc where the ALMA dust continuum reveals six dense cores, the Mach number increases up to 7. Interestingly, most of the cores are located in local minima of the Mach number. The cores are thus relatively ``calm'' compared to the surrounding turbulence. This spatial configuration has been observed in IRDCs \citep{qz2011,WangKe2012,2018A&A...611L...3S,Lu_2018}. A stark contrast is seen in cores A/B, where the mean Mach number drops from 4.5 in their surrounding to 2.8 on the cores. 
The cores themselves are still supersonic, with Mach numbers ranging 2.6-5.9 (Table. \ref{tab:core}). Only in \cite{2018A&A...611L...3S} do the cores coincide exclusively with subsonic turbulence.

This ``calm cores'' picture appears to be consistent with a dissipation of turbulence from the clump scale to core scale. However, another possibility is that the natal clump, with an extent of 0.5 pc roughly traced by VLA \nh3 (1,1) emission, is initially subsonic as a result of turbulence dissipation from even larger cloud scale ($>$1 pc, cf. \citealt{2008ApJ...672L..33W}). The low turbulence allows gravitational collapse to proceed and the clump fragments to cores, where star formation has been launched in the central part, namely cores A/B. The turbulent gas in the clump might be a feedback of the heating and energetic outflows from the forming young stars embedded in cores A/B. Because the outflow direction is roughly perpendicular to the elongated clump/filament (Fig. \ref{over}), the two ends are least affected by this feedback, and thus retain the initial turbulence and temperature.

To compare our results with other observations of massive star formation regions, we draw the histogram of pixel distribution in Mach number in Fig. \ref{mn}. The peak value of the entire G35.20 clump is at around 2.8, similar to that of the Central region. This value closely resembles to the values reported in \citep{2013MNRAS.432.3288S} (2.2) and \citep{Lu_2014} (2-3) in typical protostellar clumps, and much higher than 0.9 reported in IRDC G35.39 \citep{2018A&A...611L...3S}. Note that here only \cite{2018A&A...611L...3S} has a high spectral resolution of 0.2\kms\ as in this work (see Section. \ref{sec:1}).
IRDC G35.39 thus remains the only distant high-mass star formation region (other than Orion molecular cloud, see Section. \ref{sec:1}) where subsonic cores have been observed to dominate the whole region. G35.20, as an evolved IRDC clump, does retain the configuration of ``calm cores in turbulent gas'', but the cores as well as the clump is overall supersonic. It is of great interest to investigate turbulence at the initial stages, for example in a sample of quiescent clumps like G35.39, to place more stringent constraints on theoretical models on the role turbulence plays in high-mass star formation.

\subsection{Diverse \nh3 ortho/para ratio (OPR)} \label{sec:OPR_discuss}

The detection of all the seven rotational lines of \nh3 (1,1) to (7,7), including two ortho lines (3,3) and (6,6), enabled us to derive reliable OPR of \nh3. Many previous \nh3 studies only detected one ortho line \nh3 (3,3), or even no ortho line, and an OPR of 1 is often assumed to derive \nh3\ column density \citep[e.g.,][]{WangHierarchical,2018A&A...611L...3S}.

OPR values smaller than unity have been reported in high-mass star formation regions \citep{1985A&A...146..134H,1988A&A...201..285H,2012A&A...543A.145P}, as well as in low-mass star formation regions \citep{2022arXiv220606585F}. 
The low OPR is thought to be originated from the gas that formed \nh3 molecules \citep{2013ApJ...770L...2F}. 

On the other hand, OPR values higher than unity (1.1-3.0) have also been reported in Galactic star formation regions \citep{Goddi2011,WangHierarchical}.
In IRDC G11.11-0.12,
\cite{WangHierarchical} observed \nh3\ and find the OPR increases from 1.1 to 3.0 along molecular outflows downstream. They interpreted the trend as differential desorption of the ortho- and para-\nh3\ molecules due to a slight difference in the energy needed for desorption. In G35.20, the outflow is perpendicular to the filament/clump, but the Central/Southern regions are more affected by the outflow than the Northern region. The trend is consitent with that in G11.11-0.12, but the OPR values are much lower.

As our OPR is derived from multiple ortho and multiple para lines, which is rarely accomplished in previous studies, it should be more robust. More similar investigation in a sample of sources is needed to reach conclusive results on the \nh3\ OPR in star formation regions (Wang et al., in prep).

\section{Summary}\label{sec:6}
We have presented high-resolution imaging observations (both in terms of spatial and spectral resolution) made with ALMA and JVLA towards the massive star formation region G35.20-0.74 N. The ALMA 870\um\ dust continuum resolves fragmentation, and the JVLA NH$_3$ (1,1) to (7,7) lines and 1.2\,cm continuum emission reveal gas properties. By fitting the \nh3 line cubes, we obtain maps of kinetic temperature, excitation temperature, \nh3 column density, centroid velocity, velocity dispersion, and ortho/para ratio. Based on these results, we derive a Mach number map due to intrinsic turbulence after subtracting line broadening by channel width (10\%), thermal pressure (10\%), and velocity gradient (3\%). These well resolved maps enable a unique opportunity to study the dynamical properties of G35.20 in depth. Our main findings are summarized as follows.

\begin{enumerate}
    \item The VLA \nh3\ line emission reveals a 0.5 pc long filament/clump, an intermediate scale between the larger dust clump traced by APEX 870\um\ image and the six dense cores resolved by ALMA at the central 0.2 pc. The VLA 1.2\,cm continuum emission reveals a lane of ionized gas in agreement with a thermal jet. The jet is 37\degree misaligned to the filament traced by VLA \nh3\ and ALMA dust cores.
    \item 
    The turbulence is overall supersonic in G35.20, with the Mach number pixel distribution peaking at 2.8. However, the spatial distribution of Mach number is highly structured across the 0.5 long filament/clump: the two ends of the filament feature subsonic to transonic turbulence with Mach number down to lower than 2, the bulk of the filament has Mach number in the range of 2-4, towards the Central part of the filament it rises up to 6-7, and on the central dust cores A/B Mach number drops dramatically to $<3$.
    \item
    The central cores A/B, which are actively forming stars, appear to have decoupled from the parental filament/clump, evident by an opposite velocity gradient and reduced turbulence as compared to the filament/clump.
    \item
    The \nh3 OPR is 0.1-0.3 across the clump and has a spatial structure similar to that of \nh3 column density. Our study provides a rare case in which more than one ortho-\nh3 line is used to derive the OPR, calling for more similarly robust measurements to study the OPR in star formation regions.
\end{enumerate}

Our observations provide a snapshot of the turbulence in G35.20, an evolved IRDC clump with active ongoing star formation in its central part. G35.20 shows overall supersonic turbulence, generally consistent with the traditional picture.
However, the reduced turbulence towards the central star-forming cores indicates an initial stage with low level of turbulence, possibly similar to that of the quiescent parts of the filament with subsonic turbulence.
Importantly, because G35.20 is evolved, the turbulence is not at a pristine state. In a forthcoming paper (Wang et al. in prep), we will analyze a sample of IRDCs, in order to put more stringent constraints on the role of turbulence in the initial conditions of massive star formation.

\begin{acknowledgements}
We are grateful to the anonymous referee for constructive comments that helped improve the manuscript. 
We thank Fengwei Xu, Qizhou Zhang, Tapas Baug, Nannan Yue, Siju Zhang, Wenyu Jiao, and Yifei Ge for valuable discussion.
We acknowledge support from the National Science Foundation of China (11973013), the China Manned Space Project (CMS-CSST-2021-A09), the National Key Research and Development Program of China (22022YFA1603102, 2019YFA0405100), and the High-performance Computing Platform of Peking University.
The National Radio Astronomy Observatory is a facility of the National Science Foundation operated under cooperative agreement by Associated Universities, Inc.
This paper makes use of the following ALMA data: ADS/JAO.ALMA\#2011.1.00275.S. ALMA is a partnership of ESO (representing its member states), NSF (USA) and NINS (Japan), together with NRC (Canada), MOST and ASIAA (Taiwan), and KASI (Republic of Korea), in cooperation with the Republic of Chile. The Joint ALMA Observatory is operated by ESO, AUI/NRAO and NAOJ.

\end{acknowledgements}


\begin{appendix} 
\section{comparing one- and multi-velocity components fitting}
\label{sec:appendix}
\begin{figure*}[htb!]
    \centering
    \includegraphics[width=1.0\textwidth]{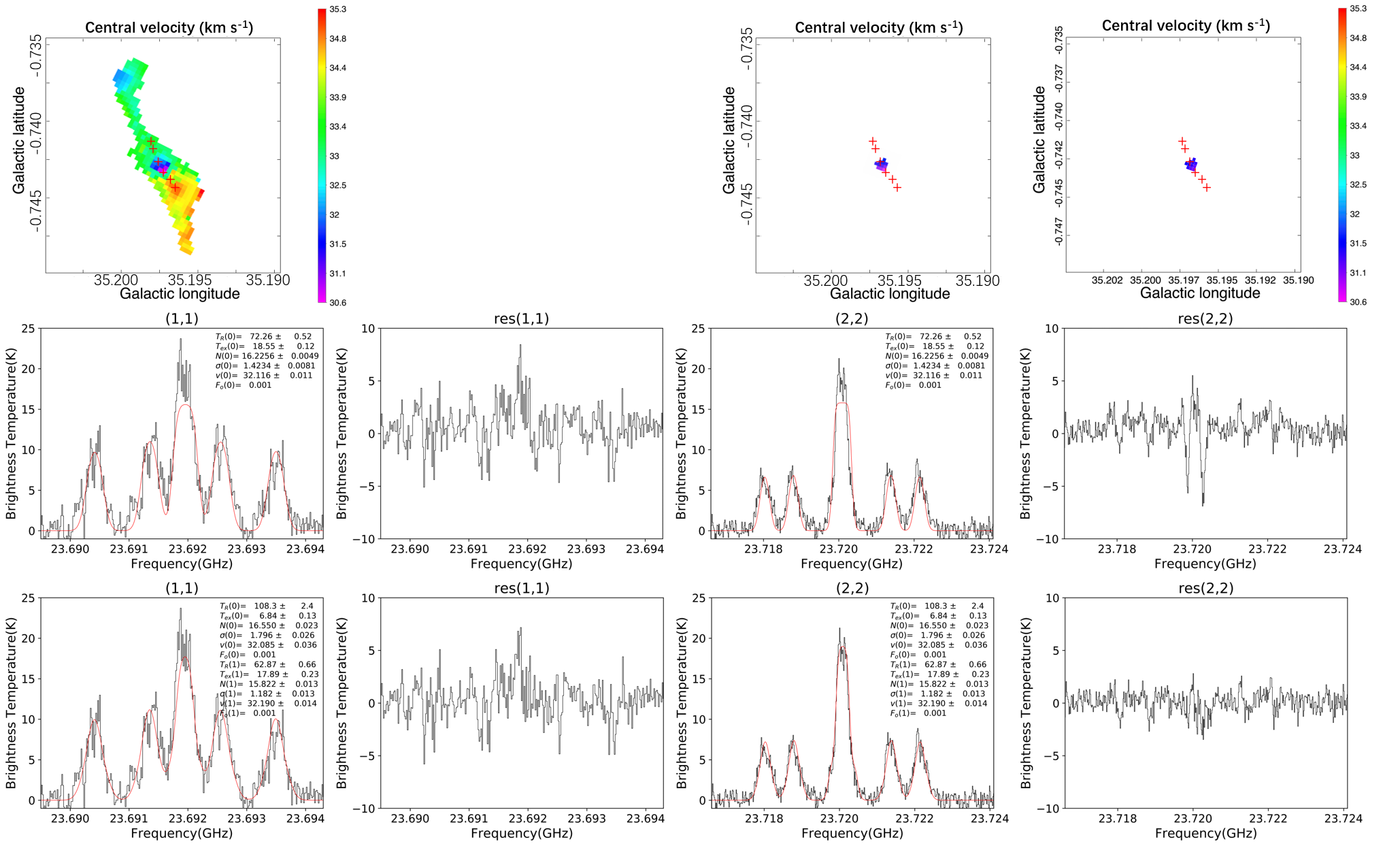}
    \caption{
Fitting results of one/two velocity components.    
    \textbf{Upper row}: 
     Fitted centroid velocity maps from the one-component fitting (left), and that of the main and secondary component from the two-component fitting (middle, right). The red crosses mark the six ALMA 870\um\ cores A-F, as in Fig. \ref{over}.
     Across the entire clump well fitted by a single component, only a tiny portion around core A can be properly fitted with two-component fitting. 
    \textbf{Middle row}:  
    One velocity component fitting. NH$_3$ (1,1) and (2,2) spectra (black) extracted from the peak of core A, overlaid with fitting results (red) and residuals. Fitted parameters are labeled on the spectra plots.
    \textbf{Bottom row}: 
    same as the middle row, but for two velocity components fitting. 
    }
    \label{fit}
\end{figure*}

The G35.20 \nh3\ clump seems to be well described by a single velocity component. In brief, fitting two velocity components failed to converge in most but 11 pixels (1.5 beam area) around core A (top row, Fig. \ref{fit}).

In the following we present the \nh3\ spectral line fitting with one velocity component, and compare the results to that of multiple velocity components fitting. The results of two velocity components fitting are presented as a representative one because other fittings including more than two velocity components are similar with that of two velocity components.

First, we run the one-component fitting on the entire clump. The middle row in Fig. \ref{fit} shows the spectra of NH$_3$ (1,1) and (2,2) extracted towards core A (marked in Fig. \ref{ve}). The observed data (black) is over-plotted with the fitted line (red) and labeled with the fitted parameters. This particular pixel is chosen to illustrate the complexity in the spectra.
Comparing the data and the best fitted spectra from the top row, we cannot exclude the existence of two or more velocity components in core A because of the two-peak structure in the residual.

Motivated by the results of the spectral fits towards core A, we then run a fitting with two components. However, only 11 pixels (about 1.5 beam area) around core A can be successfully fitted: the fitting in the vast majority of the pixels do not converge because of much worse residuals under the assumption of two components. The fitting results with two velocity components for the pixel towards core A are shown in the bottom row of Fig. \ref{fit}.

Comparing the two fits shown in Fig. \ref{fit}, the second run results in better residuals for the pixels around core A, yet the fitting is only valid in 1.5 beam area around this core, and the vast majority of the \nh3\ clump cannot be fitted. Nevertheless, in core A, the main velocity component (with $V_{\rm LSR} = 32.09\ \rm{km}\ {\rm{s}^{-1}}$) from the two-components fitting is similar to the one-component fitting (with $V_{\rm LSR} = 32.11\ \rm{km}\ {\rm{s}^{-1}}$). Thus, the centorid velocity maps are similar and both are around 32 km$\ {\rm{s}^{-1}}$  (in Fig. \ref{fit}). The Mach number derived from the one-component fitting is about 3.1 in core A, while the two-components fitting results is 3.1 (for the main component at $V_{\rm LSR} = 32.09\ \rm{km}\ {\rm{s}^{-1}}$) and 3.3 (for the secondary component at $V_{\rm LSR} = 32.19\ \rm{km}\ {\rm{s}^{-1}}$). The secondary component contributes 0.7\% to the difference of Mach number (less than the uncertainty from the fitting). 


In summary, our comparison shows that while core A can be better described with two components, there is mainly one-component in the rest of the G35.20 clump. Even in core A, the one-component fitting results can reproduce the properties of the main component in the two-component fitting. 
As an example, we show one of the six fitting parameters, centroid velocity, from the one- and two-component fitting in the top row of Fig. \ref{fit}, demonstrating the robustness of the one-component fitting.
Therefore, we adopt the one-component fitting throughout the paper.

\end{appendix}

\bibliographystyle{plainnat}
\bibliography{main}

\end{document}